\documentclass[final,a4paper,UKenglish,cleveref,autoref]{oasics-v2021}

\title{Formal verification in Solidity and Move: insights from a comparative analysis}

\author{Massimo Bartoletti}{University of Cagliari, Italy \and \url{http://blockchain.unica.it} }{bart@unica.it}{https://orcid.org/0000-0003-3796-9774}{Partially supported by project SERICS (PE00000014) and PRIN 2022 DeLiCE (F53D23009130001) under the MUR National Recovery and Resilience Plan funded by the European Union -- NextGenerationEU.}

\author{Silvia Crafa}{University of Padova, Italy}{crafa@math.unipd.it}{https://orcid.org/0000-0003-0993-4734}{Supported by the National Recovery and Resilience Plan (NRRP) Project ``Securing sOftware Platforms - SOP'', CUP H73C22000890001}

\author{Enrico Lipparini}{University of Cagliari, Italy \and \url{https://elipparini.github.io/}}{enrico.lipparini@unica.it}{https://orcid.org/0009-0009-0428-4403}{Supported by project PRIN 2022 DeLiCE (F53D23009130001) under the MUR National Recovery and Resilience Plan funded by the European Union -- NextGenerationEU.}

\authorrunning{M. Bartoletti, S. Crafa, E. Lipparini} 

\Copyright{M. Bartoletti, S. Crafa, E. Lipparini}

\ccsdesc[500]{Software and its engineering~Formal software verification}

\keywords{Smart contracts, Solidity, Move, Verification, Blockchain}

\category{} 

\relatedversion{} 

\nolinenumbers 

\newcommand{\ifempty}[3]{%
  \ifthenelse{\isempty{#1}}{#2}{#3}%
}

\newcommand{\mypar}[1]{\medskip\noindent\textbf{#1.}}

\usepackage{refcount}

\newcommand{\myfootnotetext}[1]{\footnotetext{#1\label{fn:text}%
        \edef\fnmark{\getpagerefnumber{fn:mark}}%
        \edef\fntext{\getpagerefnumber{fn:text}}%
        \ifx\fnmark\fntext\else\ClassWarning{}{footnote mark and text on different pages!}\fi}}

\newcommand{\codefont}{\fontsize{10}{10}\selectfont}
\newcommand{\code}[1]{{\tt\codefont {#1}}}

\newcommand{\Eg}{E.g.\@\xspace}
\newcommand{\eg}{e.g.\@\xspace}
\newcommand{\ie}{i.e.\@\xspace}

\DeclareMathSymbol{:}{\mathord}{operators}{"3A}

\newcommand{\nexts}{\ensuremath{\mathrm{next}}\xspace
}
\newcommand{\state}[1][]{%
    \ifthenelse{\equal{#1}{}}{\ensuremath{\mathrm{s}}}{\ensuremath{\mathrm{#1}}}%
    \xspace
}
\newcommand{\method}[1][]{%
    \ifthenelse{\equal{#1}{}}{\ensuremath{\mathrm{T}}}{\ensuremath{\mathrm{#1}}}%
    \xspace
}
\newcommand{\Prop}{\ensuremath{\mathrm{P}}}

\newcommand{\user}[1]{\emph{#1}}

\definecolor{lightcyan}{RGB}{210, 255, 255}

\newlength\myboxwidth
\definecolor{lightercyan}{rgb}{0.925, 1.0, 1.0} 
\definecolor{lesslightgray}{rgb}{0.92, 0.92, 0.92} 

\newcommand{\bartpropline}[2]{\smallskip\begin{tcolorbox}[bart]{{#1}: {``\emph{#2}''}}\end{tcolorbox}\smallskip}

\newcommand{\propline}[2]{\bartpropline{#1}{#2}}
\usepackage[utf8]{inputenc}
\usepackage{color}
\usepackage[usenames,dvipsnames]{xcolor}
\usepackage[most]{tcolorbox}

\definecolor{keywordcolor}{rgb}{0.7, 0.1, 0.1}   
\definecolor{tacticcolor}{rgb}{0.0, 0.1, 0.6}    
\definecolor{commentcolor}{rgb}{0.4, 0.4, 0.4}   
\definecolor{symbolcolor}{rgb}{0.0, 0.1, 0.6}    
\definecolor{sortcolor}{rgb}{0.1, 0.5, 0.1}      
\definecolor{attributecolor}{rgb}{0.7, 0.1, 0.1} 
\definecolor{white}{rgb}{1.0,1.0,1.0}

\definecolor{ocra}{rgb}{0.99,0.78,0.07}

\tcbset{bart/.style={
    colback=ocra!10, 
    colframe=ocra,   
    top=2pt,
    left=2pt,
    right=2pt,
    bottom=2pt,
    boxsep=1pt,
    halign=flush left,
    nobeforeafter,
    before skip=0pt,
    after skip=0pt,
    boxrule=0.3pt
}}

\usepackage{xspace}
\usepackage{nicefrac}
\usepackage{caption}
\usepackage{makecell}

\usepackage{tikz}

\usepackage[inline,shortlabels]{enumitem} 
\newlist{inlinelist}{enumerate*}{1}
\setlist*[inlinelist,1]{%
  label=(\roman*),
}

\hypersetup{
    colorlinks=true,  
    urlcolor=blue,
}

\usepackage[final,nomargin,inline,index]{fixme} 
\fxusetheme{color}
\FXRegisterAuthor{bart}{anbart}{\color{magenta} {\underline{bart}}}
\FXRegisterAuthor{enrico}{anenrico}{\color{red} {\underline{enrico}}}
\FXRegisterAuthor{silvia}{ansilvia}{\color{blue} {\underline{silvia}}}

\crefname{table}{Listing}{Listings}

\definecolor{LightGrey}{rgb}{0.975,0.975,0.975}

\lstdefinelanguage{solidity}{
, basicstyle=\ttfamily\linespread{1.15}\scriptsize\lst@ifdisplaystyle\scriptsize\fi
, commentstyle=\color{Gray}
, morecomment=[l]{//}
, morecomment=[s]{/*}{*/}
, escapechar=\$
, classoffset=0,
, keywordstyle=\color{NavyBlue}\bfseries
, morekeywords={assert,require,if,then,else,for,break,call,delegatecall,transfer,send,case, catch,continue,do,while,emit, new, return, revert, selfdestruct, try, with, throw, switch, suicide}
, classoffset=1
, keywordstyle=\color{YellowGreen}\bfseries
, morekeywords={external, implements, import, interface, internal, library, payable, pragma, private, protected, public, pure, returns, super, using, view}
, classoffset=2
, keywordstyle=\color{blue}
, morekeywords={function, constructor, contract, constant, struct, address, bool, byte, bytes, bytes1, bytes2, bytes3, bytes4, bytes5, bytes6, bytes7, bytes8, bytes9, bytes10, bytes11, bytes12, bytes13, bytes14, bytes15, bytes16, bytes17, bytes18, bytes19, bytes20, bytes21, bytes22, bytes23, bytes24, bytes25, bytes26, bytes27, bytes28, bytes29, bytes30, bytes31, bytes32, enum, int, int8, int16, int24, int32, int40, int48, int56, int64, int72, int80, int88, int96, int104, int112, int120, int128, int136, int144, int152, int160, int168, int176, int184, int192, int200, int208, int216, int224, int232, int240, int248, int256, mapping, string, uint, uint8, uint16, uint24, uint32, uint40, uint48, uint56, uint64, uint72, uint80, uint88, uint96, uint104, uint112, uint120, uint128, uint136, uint144, uint152, uint160, uint168, uint176, uint184, uint192, uint200, uint208, uint216, uint224, uint232, uint240, uint248, uint256, var, void, ether, finney, szabo, wei, days, hours, minutes, seconds, weeks, years}
, classoffset=3
, keywordstyle=\color{Plum}\bfseries
, morekeywords={balance, block, blockhash, instanceof, coinbase, difficulty, gaslimit, number, timestamp, msg, data, gas, sender, value, sig, value, now, tx, gasprice, origin}
}

\lstdefinelanguage{move}{%
, basicstyle=\ttfamily\linespread{1.15}\scriptsize\lst@ifdisplaystyle\scriptsize\fi
, commentstyle=\color{Gray},
, morecomment=[l]{//}%
, morecomment=[s]{/*}{*/}%
, moredelim=[s][{\itshape\color[rgb]{0,0,0.75}}]{\#[}{]}%
, escapechar=\$
, classoffset=0,
, keywordstyle=\color{NavyBlue}\bfseries
, morekeywords={assert!, move_to, move_from, borrow_global, borrow_mut, borrow_global_mut, exists, break, continue, else, for, if, in, loop, match, return, while}  
, classoffset=1
, keywordstyle=\color{YellowGreen}\bfseries
, morekeywords={as, const, let, move, mut, ref, static}  
, morekeywords={module, public, external, acquires}  
, keywordstyle=\color{blue}\bfseries
, classoffset=2
, morekeywords={invariant, enum, fun, impl, Self, self, struct, trait, type, union, use, where, forall, exists}
, morekeywords={signer, address, bool, char, f32, f64, i8, i16, i32, i64, isize, str, u8, u16, u32, u64, unit, usize, i128, u128}  
, classoffset=3
, keywordstyle=\color{Plum}\bfseries
, morekeywords={address_of, Err, false, None, Ok, Some, true},
, classoffset=4
, keywordstyle=\color{red}\bfseries
, morekeywords={spec, ensures,requires,post,global,spec_get}
}

\lstdefinelanguage{cvl}{
, basicstyle=\ttfamily\linespread{1.15}\scriptsize\lst@ifdisplaystyle\scriptsize\fi
, commentstyle=\color{Gray}
, morecomment=[l]{//}
, morecomment=[s]{/*}{*/}
, classoffset=0
, escapechar=\$
, morekeywords={anonymous, assembly, balance, break, call, callcode, case, catch, class, constant, continue, constructor, contract, debugger, default, delegatecall, delete, do, else, emit, event, experimental, export, external, false, finally, for, function, gas, if, implements, import, in, indexed, instanceof, interface, internal, is, length, library, log0, log1, log2, log3, log4, memory, modifier, new, payable, pragma, private, protected, public, pure, push, return, returns, revert,  selfdestruct, send, solidity, struct, suicide, super, switch, then, this, throw, transfer, true, try, typeof, using, value, view, while, with, addmod, ecrecover, keccak256, mulmod, ripemd160, sha256, sha3}
, keywordstyle=\color{NavyBlue}\bfseries
, classoffset=1
, morekeywords={storage, env, mathint, method, calldataarg, address, bool, byte, bytes, bytes1, bytes2, bytes3, bytes4, bytes5, bytes6, bytes7, bytes8, bytes9, bytes10, bytes11, bytes12, bytes13, bytes14, bytes15, bytes16, bytes17, bytes18, bytes19, bytes20, bytes21, bytes22, bytes23, bytes24, bytes25, bytes26, bytes27, bytes28, bytes29, bytes30, bytes31, bytes32, enum, int, int8, int16, int24, int32, int40, int48, int56, int64, int72, int80, int88, int96, int104, int112, int120, int128, int136, int144, int152, int160, int168, int176, int184, int192, int200, int208, int216, int224, int232, int240, int248, int256, mapping, string, uint, uint8, uint16, uint24, uint32, uint40, uint48, uint56, uint64, uint72, uint80, uint88, uint96, uint104, uint112, uint120, uint128, uint136, uint144, uint152, uint160, uint168, uint176, uint184, uint192, uint200, uint208, uint216, uint224, uint232, uint240, uint248, uint256, var, void, ether, finney, szabo, wei, days, hours, minutes, seconds, weeks, years}
, keywordstyle=\color{blue}
, classoffset=2
, morekeywords={currentContract, lastReverted, nativeBalances, selector, withRevert, lastStorage, block, blockhash, coinbase, difficulty, gaslimit, number, timestamp, msg, data, gas, sender, sig, value, now, tx, gasprice, origin}
, keywordstyle=\color{Plum}\bfseries
, classoffset=3
, morekeywords={invariant,rule,assert,satisfy,require}
, keywordstyle=\color{red}\bfseries,
}

\makeatletter
\lstdefinestyle{mystyle}{
  basicstyle=%
    \ttfamily
    \lst@ifdisplaystyle\footnotesize\fi
}
\makeatother
\newcommand{\githubaddress}{https://github.com/blockchain-unica/solidity-vs-move-verification}
\newcommand{\githuburl}{\url{\githubaddress}}
\newcommand{\sheeturl}{https://docs.google.com/spreadsheets/d/1oy43kISC2iM63XfksfonRBzjXPkpLoDglz-m24nazvc/edit?usp=sharing}

\newcommand{\specurl}[2]{\href{\githubaddress/tree/fmbc25/contracts/#1/README.md\#{#2}}{\textcolor{blue}{\texttt{#1}/\texttt{#2}}}}

\newcommand{\contracturl}[1]{\href{\githubaddress/tree/fmbc25/contracts/#1/README.md}{\textcolor{blue}{\texttt{#1}}}}

\newcounter{nBankCounter}
\newcounter{nVaultCounter}
\newcounter{nPricebetCounter}
\newcounter{nTotCounter}
\newcommand{\nTotProperties}{\arabic{nTotCounter}\xspace}

\ArticleNo{3}

\begin{document}

\maketitle

\begin{abstract}
Formal verification plays a crucial role in making smart contracts safer, being able to find bugs or to guarantee their absence, as well as checking whether the business logic is correctly implemented.
For Solidity, 
even though there already exist several mature verification tools, the semantical quirks of the language can make verification quite hard in practice. 
Move, on the other hand, has been designed with security and verification in mind, and it has been accompanied since its early stages by a formal verification tool, the Move Prover.
In this paper, we investigate
through a comparative analysis: 1) how the different designs of the two contract languages impact verification, and 2) what is the state-of-the-art of verification tools for the two languages, and how do they compare on three paradigmatic use cases.
Our investigation is supported by an open dataset of verification tasks performed in Certora and in the Aptos Move Prover. 
\end{abstract}

\section{Introduction}

Due to the immutability of the code after deployment 
and the huge amount of economic assets managed, ensuring the correctness of smart contracts is a crucial task.
Attacks exploiting code vulnerabilities and wrong implementations of the business logic are estimated to have caused  over \$6 billion of losses~\cite{Chaliasos24icse}, creating
a huge demand for safer and verifiable code. 

Solidity, the most adopted smart contract language, 
presents  semantical quirks that make contract implementation quite error-prone, and that highly complicate the verification process. 
In order to address this issue, several bug-detection tools have been developed~\cite{Tolmach22csur,Zhang23icse}, as well as some verification tools, that vary in scope, specification language, and level of abstraction. Most notably, SolCMC~\cite{Solcmc}, shipped with the Solidity compiler,
and the Certora Prover~\cite{certora}, developed for auditing.

Move is a more recent smart contract language, originally developed for the Diem/Libra blockchain  
and later adopted by Aptos, SUI and IOTA.
Designed with verifiability in mind, Move has been accompanied by a formal verification tool~\cite{Zhong20cav} since its early development.

In this work, we investigate how differences in the design of Solidity and Move (in the Aptos dialect) affect verifiability.
We base our study on a comparative analysis of a small set of paradigmatic use cases, 
each evaluated against a range of representative properties. These properties span from low-level
aspects, such as function specifications and state invariants, to more high-level ones that
characterize the business logic of the contract.
For each property, we study the ground truth in Solidity and in Move, 
and we write, whenever possible, the corresponding formal specifications in the Certora Verification Language and in the Move Specification Language.
We focus, in particular, on properties that exhibit discrepancies in ground truths, expressibility, or verifiability. 
The 
results
of our analysis offer relevant insights about 
the following  research questions: 
\begin{description}

\item[RQ1)] What is the impact of different features of  Solidity and Move on verification?

\item[RQ2)] What is the state-of-the-art of verification tools for Solidity and Move, and which kind of properties are they currently able to verify?
\end{description}
As an additional contribution, we have developed a {\bf public dataset}\footnote{\githuburl} (the first of this kind)
that serves as a basis of an experimental --- and extensible --- comparison between the Certora and the Aptos Move specification languages and verifiers.

\mypar{Structure}
The paper starts in~\Cref{sec:background} with an overview of Solidity and Move, and their respective verification tools. 
\Cref{sec:methodology} presents our methodology and discusses the choices of tools, use cases and properties. 
In~\Cref{sec:comparison} we present the results of our comparative analysis, addressing RQ1 and RQ2 in~\Cref{sec:comparison:language} and in~\Cref{sec:comparison:prover}, respectively. 
Finally, in~\Cref{sec:conclusions} we summarize our findings, discuss limitations, and outline future work.

\section{Background}
\label{sec:background}

In this section we overview the main features of 
the two languages and of the two verification tools
considered in our experimental comparison. 
In particular, we focus on key design choices of the underlying intended blockchain's model, since it has an impact on both the specification and the verification of contracts' properties.

\subsection{Contract languages: Solidity \emph{vs.}~Move}
\label{sec:background:languages}

From the perspective of smart contract programming, a blockchain is best understood as an asset-exchange state machine, in which the state keeps track of the assets owned by each account, and every transaction contributes to a state transition, possibly creating new assets or exchanging assets among accounts. 
In Solidity, there are two kinds of accounts: 
externally owned accounts (EOAs) and contract accounts.
The state of the asset-exchange machine can be seen as a map that associates each EOA with a balance of native assets owned by the account (\eg, ETH in Ethereum), and each contract account with a balance and a \emph{storage}, which contains variables and data structures that define the contract state. 
Differently, in Move the state of the machine can be seen as a map from accounts to the assets owned by them. 
Assets (called \emph{resources} in Move) are encoded by struct datatypes that enjoy linear semantics, \ie, a static type system ensures that resources are never duplicated or lost.
The main difference between Solidity and Move is the representation and accounting of  assets. 
In particular, accounts in Solidity can only explicitly own native assets, while in Move they can own arbitrary resources.
This has a relevant impact, since most real-world contracts involve the creation and exchange of \emph{user-defined} assets, \eg to represent utilities or market shares in DeFi protocols.
Representing and handling user-defined assets in Solidity requires a suitable encoding in the smart contract, while in Move all assets are dealt uniformly.

\begin{figure}[t]
\small
\begin{lstlisting}[language=solidity,caption={Simplified Solidity code for the Bank case study},label={fig:codeSol}]
contract Bank { 
  mapping (address => uint) credits;
  function deposit() payable {  credits[msg.sender] += msg.value;  }
  function withdraw(uint amount) {
    credits[msg.sender] -= amount; payable(msg.sender).transfer(amount);
  }
}
\end{lstlisting}
\vspace{-15pt}
\end{figure}

\begin{figure}[t]
\begin{lstlisting}[language=move,caption={Simplified Move code for the Bank case study},label={fig:codeMove}]
module bank {
  struct Bank { credits : SimpleMap<address, Coin> } //  resource definition 
  fun init(account : &signer) {
    let bank = Bank { credits : simple_map::new() }; // create a resource 
    move_to(account, bank); // now signer owns a bank
  }
  fun deposit(sender : &signer, owner : address, amount : u64)  {
    let bank = borrow_global_mut<Bank>(owner); // borrow the resource        
    let to_deposit = coin::withdraw(sender, amount); // get sender's coins
    let credit = map::borrow_mut(&mut bank.credits,address_of(sender)); 
    coin::merge(credit, to_deposit); // increase credit by merging coins   
  }
  fun withdraw(sender : &signer, owner : address, amount : u64) {  ... }
}
\end{lstlisting}
\end{figure}


To exemplify, consider the simplified Solidity code in~\Cref{fig:codeSol}, which encodes a simple bank contract. Once deployed on the blockchain, the global store keeps track of the bank's \emph{balance}, \ie the amount of ETH associated to the contract account. The contract stores in a variable the \code{credits} (a \emph{number} that \emph{represents} ETHs) associated to each bank's client. 
When an account (\code{msg.sender}) invokes the function \code{deposit} sending a given amount (\code{msg.value}) of ETH, the effect is twofold: the ETHs are transferred to the contract's balance, and the sender's additional credit is registered.
The  {\tt withdraw} function first decreases the number of credits and then transfers the amount of ETH to the sender. 
The corresponding Move code is shown in~\Cref{fig:codeMove}. The code defines a \code{bank} module, which relies on two types of \emph{resources}: the \code{Coin}s provided by the underlying platform, and the user-defined \code{Bank} data structure. The contract is initialized by the function \code{init}: the signer of the transaction initializes a new, empty, \code{Bank} and registers its ownership in the global store. When an account invokes the function \code{deposit}, he takes the role of the \code{sender} (\ie, the transaction's signer), and the function code executes three steps: (i) the \code{Bank} resource is borrowed from the address of its \code{owner},  (ii) an \code{amount} of \code{Coin}s are borrowed from those owned by the \code{sender}, and (iii) they are merged with those already registered in the \code{Bank}'s \code{credits} map.  

Despite its simplicity, this example already shows some differences between the two languages in terms of asset management. 
In Move, resources are first-class citizens, with properties such as linearity statically guaranteed by the type system.
By contrast, in Solidity the user-defined assets must be carefully handled at the contract level (\eg, the logic of the {\tt credits} map must correctly match the flow of ETH), which may be a source of critical bugs. 

We will discuss other features and differences between Solidity and Move in~\Cref{sec:comparison}, where they will be instrumental in addressing RQ1 about the impact of language design on smart contract verification.

\subsection{Formal verification tools for Solidity and Move}
\label{sec:background:provers}

For Solidity, there exist several bug detection tools~\cite{Kushwaha22access} as well as several verification tools~\cite{Tolmach22csur}. 
The two main verification tools are SolCMC \cite{Solcmc}, shipped with the Solidity compiler, and the Certora Prover \cite{certora}. 
Other verification tools, including
SmartACE~\cite{Wesley22vmcai}, 
SmartPulse~\cite{Stephens21sp},   
Solvent~\cite{Solvent},
VeriSolid~\cite{Nelaturu23tdsc},
VerX~\cite{Permenev20sp}, and 
Zeus~\cite{Kalra18ndss},
target various verification aspects,
each tool having its own specification language,
level of abstraction, 
and limitations.
%
In this work, we focus on Certora (see~\Cref{sec:methodology}), whose verification language (CVL)~\cite{certora-cvl} features two ways of expressing contract properties:
\emph{invariants}, which represent conditions that must remain true across contract transitions, and \emph{rules}, which are a flexible way to specify more general conditions on possible contract transitions. CVL rules can arbitrarily combine requirements on the contract state, calls to contract functions (possibly, leaving some call parameters partially specified), and assertions on the states reached upon these calls.



The Move language, since its early stages, has been tightly coupled and integrated with the Move Prover (MVP): they have been developed and maintained together, and the MVP is intended to be used routinely during smart contract development, likely to an advanced type checker.
The Move Prover specification language (MSL) \cite{aptos-msl,Xu24langsec} features different ways of expressing properties: function specifications (in terms of pre- and post- conditions), and invariants on functions, on struct datatypes, on global states, and on state transitions.
Besides bug detection tools~\cite{Song24issta}, we are only aware of  another tool that addresses formal verification of Move contracts, VeriMove \cite{VeriMove}, built upon the Solidity counter-part VeriSolid.%






As an example, consider the property ``\emph{after a successful \code{deposit}, the credits of the sender are increased exactly by  the {amount} of tokens deposited}''.  
In CVL (\Cref{fig:specCVL}), the rule first specifies the call environment \code{e} (which includes the transaction parameters \code{msg.sender} and \code{msg.value}), and stores the sender's credit before the call to \code{deposit()} in the variable \code{old\_value}. 
Then, the rule calls \code{deposit()} with environment \code{e}, and checks whether the sender's credits after the call have been increased by the amount sent.
In MSL (\Cref{fig:specMSL}), the property is expressed as a function spec (\ie, a specification targeting a single function), in terms of pre and post conditions.
The variables tagged with {\textcolor{red}{\code{post}} refer to the values of the expressions \emph{after} the call to \code{deposit}.

\vbox{%
\begin{lstlisting}[language=cvl,caption={Specification of \specurl{bank}{deposit-assets-credit} in CVL},label={fig:specCVL}]
rule deposit_assets_credit {
  env e; // environment variables of the call
  address addr_sender = e.msg.sender; // transaction sender
  mathint amount = e.msg.value; // amount of ETH tokens sent by sender to contract
  mathint old_value = currentContract.credits[addr_sender];
  deposit(e); // perform a successful call to deposit
  mathint new_value = currentContract.credits[addr_sender];
  assert new_value == old_value + amount; // verification condition
}
\end{lstlisting}}


\vbox{%
\begin{lstlisting}[language=move,caption={Specification of \specurl{bank}{deposit-assets-credit} in MSL (simplified)},label={fig:specMSL}]
spec bank_addr::bank {
  spec deposit {
    let addr_sender = signer::address_of(sender);
    let old_credits = global<Bank>(owner).credits;
    let old_value = simple_map::spec_get(old_credits, addr_sender).value;  
    let post new_credits = global<Bank>(owner).credits;
    let post new_value = simple_map::spec_get(new_credits,addr_sender).value;
    ensures new_value == old_value + amount; // verification condition
  }
}
\end{lstlisting}}
\section{Methodology}
\label{sec:methodology}

We now detail the methodology we adopted for our comparative analysis,
explaining the choices of the verification tools, use cases, and properties, and how we have built our dataset.

\subsection{Verification tools}
\label{sec:methodology:provers}

Given the variety of verification tools available, particularly for Solidity,
doing an extensive comparison of all these tools lies beyond the scope of this work.
We focus on the Certora Prover for Solidity and the Aptos MVP for Move. The choice of the Aptos MVP is straightforward, as it is, to the best of our knowledge, the only supported version of the Prover at the time of writing,
which has furthermore been used to formally verify large Move libraries, including the entire Aptos smart contract layer~\cite{Park24fmbc}.
We exclude VeriMove~\cite{VeriMove} as it only supports a strict subset of the language.
%
For Solidity, the variety of available tools is broader. 
While no single tool strictly outperforms all others in every aspect,
we choose the Certora Prover since it is the tool most used in real-world settings for the verification of complex properties.
We will nonetheless explicitly mention other tools capable of 
addressing 
properties beyond the scope of the two selected tools, whenever applicable. 
In the following, we will refer to the two tools just as Certora and Move Prover (or MVP).
We remark that Certora and MVP have been designed with different goals. 
In Move, specification and development go side-by-side. 
Certora, on the other hand, is more oriented to the ex-post analysis of contracts and is primarily used for auditing~\cite{certora-reports}:
consequently, CVL is designed to support the verification of complex properties without requiring modifications to the contract code (\eg, updating ghost variables at given program points).
Despite these differences, applying the state-of-the-art tools to a common benchmark is crucial to answer our research questions, namely which properties can be verified in the two languages at the time of writing (RQ2), and how the choice of the contract language affects the quality of the verification process (RQ1).

\subsection{Use cases}
\label{sec:methodology:contracts}

In the selection of the verification use cases, 
we identify three paradigmatic smart contracts with increasing level of complexity and exhibiting a rich spectrum of features: a \contracturl{bank} contract (already described in~\Cref{sec:background:languages}), a \contracturl{vault} contract, and a \contracturl{price-bet} contract.

The \contracturl{vault} contract implements a security mechanism to prevent an adversary who has stolen the owner's private key from stealing their tokens.
Upon creation, the owner specifies its private key, a recovery key, and a wait time.
The contract has the following entry points:
\begin{itemize}

\item \code{receive(amount)}, which allows anyone to deposit tokens into the contract;

\item \code{withdraw(receiver, amount)}, which allows the owner to issue a withdraw request, specifying the receiver and the desired amount;

\item \code{finalize()}, which allows the owner to finalize the pending withdraw after the wait time has passed since the request;

\item \code{cancel()}, which allows the owner of the recovery key to cancel the pending withdraw request during the wait time.
\end{itemize}


The \contracturl{price-bet} contract implements a bet on a future exchange rate between two tokens.
To create the contract, the owner specifies:
itself as the contract owner;
the initial pot, which is transferred from the owner to the contract;
an oracle, \ie a contract that is queried for the exchange rate between two given tokens;
a deadline; 
a target exchange rate, which must be reached in order for the player to win the bet.
The contract has the following entry points:
\begin{itemize}
\item \code{join()}, which allows a player to join the bet. This requires the player to deposit an amount of native cryptocurrency equal to the initial pot;
\item \code{win()}, which allows the player to withdraw the whole contract balance if the oracle exchange rate is greater than the bet rate. 
This action is disabled after the deadline;
\item \code{timeout()}, which can be called by anyone after the deadline, and transfers the whole contract balance to the owner.
\end{itemize}
%

We implement each use case in Solidity and in Aptos Move, ensuring that these implementations remain as close as possible.
The verification of these use cases requires to deal with properties featuring several aspects, such as: key-value maps, access control, time constraints, contract-to-contract calls, and transaction-ordering dependencies.

\subsection{Properties }
\label{sec:methodology:props}

For every use cases, we consider an extensive set of properties, ranging from low-level properties that only target single contract functions, to more high-level ones that characterize the global behaviour of the contract. 
Our choice of properties is based on breadth and diversity (in terms of language features involved, abstraction level, temporal logic structure). 
Overall, we end up with \nTotProperties properties. 
Aiming at generality, and potentially including also properties that cannot  be expressed in the considered tools, we write properties in natural language. 
We then encode each property, whenever possible, as CVL and MSL specifications.
Often, this translation involves adding suitable low-level technical assumptions, to make the specification aligned with the spirit of the corresponding natural language property. 
As an example, in Move, users may have a frozen coin store that prevents them from receiving tokens; in such a case, even if the natural language property does not mention such aspect, we consider adding such low-level technical assumptions as part of the translation process. 
Furthermore, coherently with most verification tools,
we neglect transaction fees.
We then manually annotate the expected truth value in Solidity and Move.
Finally, we run the provers and take note of their output.
We end up, for each use case, with a \href{\sheeturl}{sheet} consisting of four main columns for each property row: two columns for the ground truths, and two for the provers results. We enrich the table with additional columns containing notes on: the class of the property, the expected truth values, the formal specifications, and the provers outputs.

\section{Comparison}
\label{sec:comparison}

Based on our dataset, we now present our comparative analysis.
Building upon the analysis of each property, we elaborate our findings to construct an organized knowledge that extends beyond our choice of use cases. 
In particular, we focus on properties where discrepancies arise between verification in Solidity and Move.
These properties serve as illustrative examples for a broader discussion of the fundamental differences in the verification of the two languages.
Our observations can be grouped as follows: 
properties whose ground truths disagree;  
properties that trivially hold in one language but not in the other; 
properties not expressible in one or both specification languages; 
properties expressible but not verifiable by one or both tools.
These four cases are not necessarily independent of one another, but they help to better identify the primary causes of discrepancy. 
In the first two groups, the discrepancy specifically depends on the contract languages, while, in the latter two, it depends more on the specification language and prover functionalities. 
We accordingly organise this part into two subsection: \Cref{sec:comparison:language} focuses on the impact of the contract languages, while \Cref{sec:comparison:prover} focuses on the impact of the specification languages and on the provers functionalities.

\subsection{Impact of the contract language}

\label{sec:comparison:language}

\mypar{Resource preservation} 
As observed in~\Cref{sec:background}, 
Move enforces asset integrity by ensuring that assets cannot be duplicated
but only \emph{moved} between owners;
by contrast, Solidity --- except that for native tokens (ETH) --- requires the management of assets to be implemented at a contract level.
For example, in the \contracturl{bank} use case, 
the \code{credits} are rendered in Move as a map from \code{address} to \code{Coin} (that \emph{are} the actual assets), 
while in Solidity they are
a map from \code{address} to \code{int}.
This means that the Solidity code merely \emph{tracks} the assets deposited by each user. However, implementation bugs can lead to a mismatch between the assets controlled by the contract and the overall amount of user credits, assigning more or fewer credits than they are entitled to. 
This significantly impacts the specification and verification of properties.
First of all, in MSL, since credits \emph{are}  assets, such properties are implied by properties that concern assets. 
For example, in MSL the specification  of the property 

\propline{\specurl{bank}{deposit-assets-credit}}{after a successful deposit of $n$ tokens, the credits of the sender are increased by $n$}
\noindent
is exactly a sub-specification of the property 

\propline{\specurl{bank}{deposit-assets-transfer}}{after a successful deposit of $n$ tokens, $n$ tokens pass from the control of the sender to the control of the contract.}

In CVL, by contrast, these two properties are disjoint, and it is possible --- in the presence of bugs related to the handling of credits --- for the former to hold while the latter is violated.
This shows that, to cover the same set of properties, Solidity requires a greater number of specifications than Move.
Moreover, in Move, certain properties concerning credits trivially hold, while, in Solidity, they may be hardly verifiable, or even unexpressible.
For example:  
\propline{\specurl{bank}{credits-leq-balance}}{the assets controlled by the contract are (at least) equal to the sum of all the credits}
\noindent
trivially holds in Move, where \code{credits} coincide with the 
deposited 
assets, 
but not in Solidity, where \code{credits} just {represent}  the deposited assets. 
In general, verifying such kind of properties is quite challenging, as they require to reason about quantities depending on an unbounded number of users.

\mypar{Access control and ownership} 
Most smart contracts implement access control mechanisms to ensure that certain actions can only be performed by certain users under certain conditions.
A typical check is that some resources can only be updated by functions called by the contract owner.
Move inherently supports this kind of check: it suffices that all the functions that update the resource borrow it through a signer address. This is because, in Move, a resource can only be referenced through the address of its resource owner. %
This is a security pattern in Move to reduce the risks of access control errors~\cite{aptos-security-guidelines}.
%
In Solidity, instead, resource ownership is not a native notion, so it must be encoded by the contract logic. In particular, in order to implement the check above, the contract must first record the owner address in a variable, and each sensitive function must require that the transaction sender and the owner coincide. 
Forgetting even a single check can lead to vulnerabilities, as in the Parity Wallet hack, where the absence of such check in a function enabled the attacker to become the owner and steal all the contract funds~\cite{parity-wallet-hack}. 
In our dataset this difference in behaviour can be observed, \eg, for the property 
\propline{\specurl{vault}{finalize-revert}}{a call to \code{finalize()} aborts if the sender is not the owner.}
In CVL, we need to explicitly check that the address of the sender is equal to the \code{owner} field, while, in MSL, since the function directly accesses the \code{Vault} struct owned by the sender, and the \emph{owner} is not determined by the value of a variable 
but by the address that owns the resource,
then the property trivially holds, being enforced by the language.
Another typical check is that some addresses used by the contract (\eg, its owner) do not change throughout the contract lifespan (\eg, \specurl{vault}{owner-immutable}).
In Solidity, it is possible to enforce that by declaring the addresses as \code{immutable}. 
In such a case, the property is directly enforced by the Solidity compiler, without having to resort to verification. 
Enforcing the same check in Move is less straightforward. A method is to record the concerned addresses as fields of some struct, and then verify with the MVP that these fields are invariant.

\mypar{Assets transfer} \label{sec:comparison:language:assetTransfer}
Solidity and Move render assets and their transfers differently, leading to different techniques for expressing and verifying properties related to them.
In Solidity, while there is a clear dichotomy in how the native asset (\ie, ETH) and user-defined assets (\eg, ERC20 tokens) are handled, in both cases transfers are rendered as contract calls.
The outcome of a contract call depends on whether the callee is an externally owned account (EOA) or a contract account.
When the callee is an EOA, the transfer is guaranteed to succeed, whereas for contract accounts the effect of the call depends on the implementation of the function handling the call.
For instance, assets may be returned to the caller if the call reverts, 
or they may be forwarded (either in full or in part) to other accounts if the function is designed to do so and has enough gas.
Therefore, properties about asset transfers should either discriminate between EOAs and contract accounts, 
or add assumptions about the implementation of the receiver function.
However, the first choice is not always viable, as detecting whether an address is an EOA or a contract account (either at contract or specification level) is possible only in limited cases~\cite{openzeppelin-eoa,BFMPS24fmbc}.
The second choice is problematic as well, since if the assumptions are false then the property may be violated at runtime.
Unlike Solidity, Move offers linguistic primitives for transferring ownership of resources, enabling a more disciplined modelling of asset transfers. 
This reduces the effort required to incorporate the necessary assumptions when encoding properties.
In our dataset, we have observed this, \eg, in the property 
\propline{\specurl{vault}{finalize-not-revert}}{a \code{finalize()} transaction sent by the contract owner, in state REQ, and after  the wait time has passed, does not abort}
\noindent
which holds in Move but not in Solidity, since the transfer may fail when the receiver is a contract.
Furthermore, also 

\propline{\specurl{vault}{finalize-assets-transfer}}{after a successful \code{finalize()}, a given amount of assets pass from the contract to the \code{receiver}}
\noindent
holds in Move but not in Solidity since, if the receiver is a contract, the assets can immediately be transferred to another address through the fallback function.

\mypar{Function dispatching}
Solidity features a form of dynamic dispatching, in that the compiler does not always know, for a contract-to-contract call, the code that will be executed in the callee. 
This poses significant challenges to verification. 
Indeed, to avoid unsoundness, verification tools must assume that contract-to-contract calls can execute arbitrary code, which easily leads to false negatives.
In order to address the issue, Certora 
allows users to specify a set of possible implementations of the callee, and verify the caller against each of them~\cite{certora-dispatcher}.
This technique can require considerable effort, and does not resolve the underlying unsoundness issue.
Move, on the other hand, features static dispatching, \ie the compiler (and, consequently, the verifier) know exactly the code that will be executed in the callee. In particular, Move does not support inheritance nor any form of
method redefinition.
We have observed the impact of these different dispatching designs, \eg, in 
\propline{\specurl{price-bet}{win-revert}}{a \code{win} transaction aborts if the  oracle exchange rate is smaller than the bet exchange rate.}
%
In Certora, verifying the property requires the user to explicitly instruct the verifier to resolve the call with a given oracle implementation: leaving that unspecified would make verification fail.
In practice, many Solidity contracts are written in a way that makes it impossible to predict the actual implementations of the callees (\eg, Solidity contracts using ERC20-compatible tokens usually define only their interface).

\mypar{Other features}
\emph{Immutability.} In Solidity, the \code{immutable} keyword allows to enforce that certain variables cannot change value throughout the whole lifespan of the contract, making certain properties (\eg, the above-mentioned \specurl{vault}{owner-immutable}) enforced by the Solidity compiler. In Aptos Move, since an equivalent modifier does not seem to be available, such properties have to be explicitly verified with a prover.\footnote{Note that, in SUI Move, 
it is possible to define \emph{frozen objects} (\ie objects that cannot be modified nor moved). It does not seem possible to define \emph{frozen fields} of an object, though.}

\emph{Self-destruct.} 
In Solidity, contracts can receive native tokens at any time through the \emph{self-destruct} method. This requires additional precautions during implementation to prevent funds from getting locked in the contract.
For example, our Solidity implementation of the \contracturl{bank} use case allows users to withdraw only the funds corresponding to their credits (\ie, funds that have been previously deposited). In contrast, funds received via self-destruct cannot be withdrawn from the contract and remain locked.
This is not the case in Move, as no equivalent of the self-destruct method exists.
For example, the property 
\propline{\specurl{bank}{no-frozen-assets}}{if the contract controls some assets, then it is always possible to transfer them  to some user}
\noindent
holds in Move, but not in Solidity, since the contract only allows creditor to withdraw the assets they have deposited, but  does not  provide any function to transfer funds received via self-destruct, resulting in funds getting stuck in the contract.

\mypar{Necessary technical assumptions} As discussed in~\Cref{sec:methodology},  the translation of properties written in natural language to formal specification often requires the addition of low-level technical assumptions. Here, we report the cases that we have observed in our experiments.

\emph{Accepting incoming transfers.} As observed in the  ``\emph{Assets transfer}'' paragraph, properties concerning the transfer of assets may need further assumptions on the receiver. In Move, the only technical assumption we had to add in our dataset is that the \code{CoinStore} of the receiving address is not \code{frozen}.
In Solidity, one sufficient condition that can be used when the receiver equals to the transaction sender is that the sender is an EOA.
Although this could be encoded in CVL by requiring that 
\code{e.msg.sender==e.tx.origin}, the Certora prover does not use this additional assumption, leading to a false negative.
This is the case, \eg, of:
\propline{\specurl{bank}{withdraw-assets-transfer}}{after a successful \code{withdraw(amount)}, exactly \code{amount} units of T pass from the control of the contract to that of the sender}
\noindent
Other conditions, such as ensuring that the receiver does not fail or does not perform further calls, do not appear to be expressible in CVL.

\emph{Coin-to-FungibleAsset.} Aptos has recently introduced a ``\emph{Fungible Asset}'' (FA) standard~\cite{aptos-fungible-asset} that extends the \emph{Coin} standard, enabling automatic migration from Coin to FA by default. This automatic migration can make certain properties  concerning the transfer of Coins  violated, since Coins are not preserved (but migrated to FA). This is the case, \eg, of:
\propline{\specurl{bank}{deposit-revert}}{a transaction 
 \code{deposit(amount)} aborts if \code{amount} is greater than the T balance of the transaction sender.}
\noindent
In order to verify such properties, it is necessary to disable the automatic migration.      

\emph{Sender is not the contract.} In Solidity, it is possible that a contract calls itself. 
In certain cases, it may be necessary to assume that this is not the case, as, otherwise, certain properties might either not hold, or be unverifiable in practice. For example, \specurl{bank}{deposit-assets-transfer} specifies that, after a successfull deposit of $n$ tokens, the balance of the sender is decreased by $n$. 
While this property is true without further assumptions (since the specific \code{Bank} contract cannot call itself), in Certora the verification will fail without adding the assumption that the sender is not the contract.
This is because verification tools usually over-approximate the set of possible executions, thus considering also the impossible case in which the contract calls itself.

\subsection{Impact of the specification language and prover functionalities}
\label{sec:comparison:prover}

We now consider different classes of properties and discuss how (and whether) they can be expressed in the two specification languages. 
The organization in classes has not to be intended as a formal taxonomy,
rather as a schematic way to present our findings.

\mypar{Function specs} We denote by ``{function spec}'' properties that specifically target  a given function.
We divide these properties into ``{success conditions}'', which characterize the conditions under which a function aborts or not, and ``{post-conditions}'', which express properties regarding the state after the call, assuming that the call has not aborted. 
The Move Prover has an ad-hoc specification format for function specs. 
In Certora, function specs can be expressed as rules that explicitly mention the function being called, 
and using \code{requires} statements for pre-conditions, the expression \code{lastReverted} for checking abort conditions, and the statement \code{assert} for post-conditions. 
\Cref{fig:specCVL} and \Cref{fig:specMSL} presented in \Cref{sec:background:provers} are examples of function specs in CVL and MSL, respectively.
Both tools perform well over properties of this kind in our dataset. 

\mypar{State invariants} We denote by ``state invariants'' properties of the form ``\emph{for every reachable state \state, it holds that $\Prop(\state)$}'', where {$\Prop(\state)$} is a property that only mentions variables in the state \state. 
In Move,  state invariants can be proved in two ways: either using a \emph{struct invariant} spec, in case an invariant only deals with a single structure (\eg, in any state, the vault state is IDLE or REQ, \ie \specurl{vault}{state-idle-req-inter}), or, otherwise, using a \emph{global invariant} spec (\eg, the owner and the recovery keys are distinct, \ie \specurl{vault}{keys-distinct}).
In Certora, 
there is a common way to write invariants. 
Both tools perform well over properties of this kind on our dataset.

\begin{lstlisting}[language=cvl,caption={Specification of \specurl{vault}{state-idle-req-inter} in CVL}]
invariant state_idle_req_inter()
 currentContract.state == Vault.States.IDLE || currentContract.state == Vault.States.REQ;
\end{lstlisting}

\begin{lstlisting}[language=move,caption={Specification of \specurl{vault}{state-idle-req-inter} in MSL as struct invariant}]
spec vault_addr::vault { spec Vault { invariant (state == IDLE) || (state == REQ); } }
\end{lstlisting}

\mypar{Single-transition invariants} 
We denote by ``single-transition invariants'' properties of the form ``\emph{for every reachable state \state, and for every transaction \method, either $\method$ aborts, or it holds that $\Prop(\state, \nexts(\state, \method), \method)$}'',
 where $\nexts(\state, \method)$ is the state after a successful execution of \method in \state.  
 Note that function specs are a special case where the called function is fixed.
Certora is quite flexible for the verification of such properties, and allows to express arbitrary (quantifier-free) conditions on the parameters of \method.
In the Move Prover, there are two different ways to express single-transition invariants, both of which are less general than Certora rules.
The first way is to use \emph{global invariant updates}. 
This construct, however, does not 
allow to make explicit mention of the parameters of the transaction \method,   restricting expressible properties to those of the form $\Prop(\state, \nexts(\state, \method))$, where \method remains implicitly universally quantified. 
The second way is to use a \emph{schema} of function specs (that is, syntactic sugar to group together a set of function specs with a common body). 
Writing a single-transition invariant this way, however, requires to write an instance of the schema for each method,
making the MSL spec significantly more verbose than in CVL.
As an example, consider the property:
\propline{\specurl{bank}{assets-dec-onlyif-deposit}}{if the assets of a user \user{A} are decreased after a transaction, then that transaction must be a \code{deposit()} where \user{A} is the sender}
\noindent
In CVL, it is possible to succinctly express such property as follows: 

\vbox{%
\begin{lstlisting}[language=cvl,caption={Specification of \specurl{bank}{assets-dec-onlyif-deposit} in CVL},label={fig:specCVLsingleTransInv}]
rule assets_dec_onlyif_deposit {
  env e; method f; calldataarg args; address a;
  require e.msg.sender != currentContract && a != currentContract; 
  
  mathint old_a_balance = nativeBalances[a];
  f(e, args); // non-reverting call to an arbitrary function f of the Bank contract
  mathint new_a_balance = nativeBalances[a];
  
  assert new_a_bal < old_a_bal =>  // if the balance has decreased...
    (f.selector == sig:deposit().selector && e.msg.sender == a); // ...then f=deposit 
}
\end{lstlisting}}

In MSL, it is only be  possible to specify the contrapositive, i.e. that, for every transaction that is not a \code{deposit()}, or for which \user{A} is not the sender, then the assets of \user{A} are not decreased.
This, however, requires to write a spec for each function except \code{deposit()}, and one further function spec for the \code{deposit()}, restricted to the case of \user{A} not being the sender.

\vbox{%
\begin{lstlisting}[language=move,caption={Specification of \specurl{bank}{assets-dec-onlyif-deposit} in MSL}]
spec bank_addr::bank {
  spec withdraw {
    let a = signer::address_of(sender);
    let old_a_bal = global<coin::CoinStore<AptosCoin>>(a).coin.value;
    let post new_a_bal = global<coin::CoinStore<AptosCoin>>(a).coin.value;
    requires !features::spec_is_enabled(features::COIN_TO_FUNGIBLE_ASSET_MIGRATION);
    ensures new_a_bal >= old_a_bal;
  }
  spec deposit {
    let a = signer::address_of(sender);
    ensures forall b: address where b!=a : // b is not the sender 
             global<coin::CoinStore<AptosCoin>>(b).coin.value 
      >= old(global<coin::CoinStore<AptosCoin>>(b).coin.value);
  }
}
\end{lstlisting}
}

Note that, in the case \contracturl{bank} had a greater number of functions, the size of the MSL specification would grow proportionally, while the CVL spec size would remain constant.

\mypar{Multiple transition invariants} 
We denote by ``multiple-transition invariants'' properties of the form 
``\emph{for every reachable state \state, and for every sequence of transactions $\vec{T} = \method_1 \ldots \method_n$, either one transaction aborts, or 
$\Prop(\state, \nexts(\state,\vec{T}[1:1]), \dots , \nexts(\state,\vec{T}[1:n]), \method_1 \ldots \method_n)$} holds'', 
where $\nexts(\state,\vec{T}[1:i])$ 
denotes the  state after the successful execution of $\method_1, \dots, \method_i$.
In CVL, it is possible to express such specifications analogously to single-transition invariants, by subsequent function calls in the same rule. 
In MSL, this kind of specifications does not seem to be expressible. 
For example, consider the property: 
\propline{\specurl{vault}{finalize-or-cancel-twice-revert}}{a \code{finalize()} or a \code{cancel()} transaction aborts if performed immediately after another \code{finalize()} or \code{cancel()} transaction.}
\noindent
This is not expressible in MSL, while Certora can verify the following CVL spec: 

\vbox{%
\begin{lstlisting}[language=cvl,caption={Specification of \specurl{vault}{finalize-or-cancel-twice-revert} in CVL}, label={fig:specCVLmultipleTransInv}]
rule finalize_or_cancel_twice_revert {
    env e1, e2; bool b1, b2; // environments and selectors for transactions tx1,tx2
    if (b1) { finalize(e1); } else { cancel(e1); } // tx1 performs finalize of cancel
    if (b2) { finalize@withrevert(e2); } else { cancel@withrevert(e2); } // same for tx2  
    assert lastReverted; // checks that the 2nd tx is always reverted
}
\end{lstlisting}}

\mypar{Metamorphic properties} 
These are properties that involve multiple finite sequences of transactions~\cite{Chen18csur}.
A typical class of metamorphic property are \emph{additivity properties}: \eg: 
\propline{\specurl{bank}{withdraw-additivity}}{two successful \code{withdraw()} of $n_1$ and $n_2$ units of token T performed by the same sender are equivalent to a single \code{withdraw()} of $n_1+n_2$ units of T}
\noindent
In CVL, it is possible to express some metamorphic properties  through the use of \code{storage} types, 
which allow to record the contract storage at different points of execution and to later compare them (see, \eg \Cref{fig:specCVLmetamorphic}). 
This feature is not present in MSL, so metamorphic properties do not seem expressible.

\begin{lstlisting}[language=cvl,caption={Specification of \specurl{bank}{withdraw-additivity} in CVL (simplified)}, label={fig:specCVLmetamorphic}]
using Bank as c;
rule withdraw_additivity {
    env  e1, e2, e3; // environments for transactions tx1,tx2,tx3
    uint v1, v2, v3; // values sent along with transactions tx1,tx2,tx3  
    storage initial = lastStorage; // save the current storage in variable initial 

    require e1.msg.sender == e2.msg.sender; // the senders of tx1,tx2 must be equal
    require v1+v2 <= currentContract.opLimit;
    withdraw(e1,v1); withdraw(e2,v2); // perform tx1,tx2 in sequence
    storage s12 = lastStorage; // saves the current storage in variable s12

    require e3.msg.sender == e1.msg.sender; // the sender of tx3 is the same as tx1,tx2
    require v3 == v1+v2; // the amount of tx3 must be the sum of the amounts in tx1,tx2
    withdraw(e3,v3) at initial;
    storage s3 = lastStorage; // saves the current storage in variable s3

    assert s12[c] == s3[c];   // checks that tx1;tx2 have the same effect of tx3   
}
\end{lstlisting}
    

\mypar{Other properties}
Some classes of properties do not seem expressible in any of the two tools. 
Without claiming exhaustivity, we now briefly discuss some of the classes we have encountered, with particular attention to those that seem addressable by other tools.

 \emph{Liveness.} Liveness properties have the form 
 ``\emph{eventually a state that satisfies certain conditions is reached}''.
In \contracturl{price-bet}, a desirable liveness property is 
\propline{\specurl{price-bet}{eventually-balance-zero}}{eventually the contract balance goes to $0$}
\noindent Note that this property is closely related to, but more abstract than, the property
\propline{\specurl{price-bet}{timeout-not-revert}}{a transaction \code{timeout()} [which transfers the assets controlled by the contract to the owner] does not revert if the deadline has passed}
\noindent Tools able to handle such kind of properties, usually under the assumption of fairness conditions (in the example, that the \code{timeout()} function is called at least once after the deadline), are VeriSolid~\cite{VeriSolid}, VeriMove~\cite{VeriMove}, and SmartPulse~\cite{Stephens21sp}.

\emph{Liquidity/Enabledness.} Liquidity~\cite{Solvent} or Enabledness~\cite{Schiffl24fmbc}  properties are of the form 
``\emph{in every reachable state, certain users are always able to 
fire a (fixed) number of transactions
to reach a desirable state}''. 
In \contracturl{bank}, an example of such properties is
\propline{\specurl{bank}{no-frozen-credits}}{if the credits are strictly positive, it is possible to reduce them}
\noindent Note that this kind of property never mentions the function that should be called nor its parameters, as they  are existentially quantified and determining them (as a function of the current state)
is a task of the tool. A tool that addresses such kind of properties is Solvent~\cite{Solvent}.


\emph{CTL fragment}: The specification language of VeriSolid (and, consequently, of VeriMove) covers an expressive fragment of Computational Tree Logic (CTL). Such expressivity comes at the expenses of soundness, as the verification process relies on a certain level of abstraction.
Examples of CTL specifications include the Liveness seen before, as well as properties of the form ``\emph{$\Prop_1$ cannot happen  after $\Prop_2$}'', or ``\emph{If $\Prop_1$ happens, then $\Prop_2$ can only happen after $\Prop_3$ happens}''. 
These properties cannot be expressed in CVL, since it is not possible to talk about unbounded sequences of method calls, but only about sequences of states of finite length.
\Eg, a property not expressible in CVL but in the CTL fragment supported by VeriSolid is:
\propline{\specurl{vault}{finalize-after-withdraw-not-revert}}{after a successful \code{withdraw()}, if no \code{cancel()} or \code{finalize()} have been called successfully, then \code{finalize()} does not abort}}
\noindent
All these properties have a higher level of abstraction  than those discussed in the previous paragraphs.
Although some of these properties, in certain cases,
can be reformulated in terms of more concrete properties that imply them,   
doing so requires a more advanced knowledge of the low-level aspects, and reduces their generality.
It has been observed that properties that abstract the system 
have a better return-on-investment than low-level properties~\cite{Xu24langsec}. 

\mypar{Orthogonal features of properties} We finally address specific features of properties that can appear in all previous classes, hence for which a separate discussion is needed.

\emph{Inter vs. Intra function invariants}  
Invariants can be of two kinds: those that must be preserved across function calls (\emph{inter-function} invariants) and those that must be preserved within the execution of a function (\emph{intra-function} invariants).
In the latter, the notion of \emph{reachable state} is extended to intermediate states. 
In some cases, intra-function invariants give stronger security guarantees. 
For example, consider the invariant
\propline{\specurl{vault}{keys-invariant-inter}}{the \code{receiver} key cannot be changed after initialization}
\noindent
Requiring this invariant to only hold inter-function is not enough, as it does not capture attacks where an adversary 
\begin{inlinelist} 
\item changes the \code{receiver} key within \code{finalize()} before the transfer, 
\item sends the contract tokens to her address, and \item restores the key to the original value before the end of the function. 
\end{inlinelist} 
It is necessary to require the invariant to hold also intra-function (\specurl{vault}{keys-invariant-intra}).
In Certora, verification of intra-function invariants  is possible through ghost variables and hooks~\cite{certora-hooks}.
In Move, on the contrary, verification of intra-function invariants is, in general, not possible. The MVP can check that an invariant holds \emph{globally}, \ie every time the global state is updated~\cite{aptos-msl}, but this cannot capture every change that occurs during the execution of a function.
In the example attack mentioned above, the MVP is not able to detect that that the \code{receiver} key is changed within the execution of the \code{finalize()}.

\emph{Nested quantifiers.} Several interesting properties require the nesting of quantifiers. 
In Certora, quantifier nesting is limited to \emph{exists-forall} fragments, whereas \emph{forall-exists} fragments are disallowed. 
This makes not expressible in CVL properties such as: 
\propline{\specurl{bank}{exists-at-least-one-credit-change}}{after a successful transaction, the credits of at least one account have changed}  
\noindent 
Although MSL allows arbitrary combinations of quantifiers, in practice the verification of such properties can be problematic, as the underlying SMT solvers often struggle with quantifiers. 
In our experiments, we managed to successfully verify the previous property,
but got an inconsistent result in the case of \specurl{bank}{exists-unique-asset-change}. 
This inconsistency may be caused by 
the version of the underlying SMT solver used.

\emph{Gas.} 
As discussed in ``\emph{Assets transfer}'' in~\Cref{sec:comparison:language:assetTransfer},
the truth of certain properties may depend on the amount of gas available to the involved functions. 
For instance, in Solidity the ground truths of
\specurl{bank}{withdraw-assets-transfer} and
\specurl{vault}{finalize-assets-transfer} differ because 
of the functions used in the respective contracts to transfer ETH
from the contract to another address:
in the implementation of \contracturl{bank}, we are using \code{transfer}, 
which do not carry enough gas to perform further calls,
while in \contracturl{vault} we are using \code{call}, which
instead transfers all the gas to the callee.
Certora however over-approximates the amount of gas available, so it gives a false negative for \specurl{bank}{withdraw-assets-transfer}.

\section{Conclusions}
\label{sec:conclusions}

The empirical analysis of our study validates the folklore knowledge 
that Move is better suited for verification than Solidity.
In particular, Move's resource-orientation facilitates the verification of properties concerning, \eg, resource preservation, ownership, and transferring of assets.
The only weak spot we have observed in (Aptos) Move
is the lack of a construct to enforce the immutability of contract variables ---
a feature that instead is present in Solidity.
We have noted that, in order to properly specify certain properties and determine their truth, some low-level aspects of the underlying contract layers must be taken into account.
While this could be discouraging for smart contract developers unfamiliar with these low-level details, it can also serve as an incentive to deepen their understanding on these aspects, ultimately leading to more secure smart contract implementations.  

Concerning verification tools,
we have observed that the Certora Prover can express a broader
set of properties than the Move Prover, \eg, 
transition invariants involving multiple transactions,
metamorphic properties,   
and intra-function invariants.
We believe that all the functionalities needed to verify such properties 
could be smoothly added to the Move Prover, as well. 
We have also noted that there are several relevant classes of properties that are out of the scope of both tools (\eg, liveness, liquidity/enabledness, and, more generally, other complex temporal properties concerning the business logic of the contract).
We have observed that some of these properties can be addressed by other tools, although their current maturity level remains below that of the Certora and Move provers.

We have contributed with an open dataset of smart contract implementations and verification tasks performed in the two tools (the first of this kind), that we envision will further encourage research on 
formal verification of Solidity and Move.

\mypar{Limitations}
Although our empirical analysis is based on a set of \nTotProperties verification tasks covering a broad range of properties, we expect that extending our dataset would highlight additional differences between verification in Solidity and Move. Moreover, it could reveal some further kinds of properties that would be desirable to verify on real-world smart contracts but currently fall beyond the scope of existing verification tools.
This could be the case, \eg, of economic properties of DeFi protocols, whose verification currently requires either using weaker analysis techniques than formal verification (\eg, property-based testing~\cite{Milo22fmbc}, statistical model checking~\cite{BartolettiCJLMV22isola}), taint analysis~\cite{Wang19pacmpl,Kong23defitainter}, type systems~\cite{Yao24scif,Zhang24icse}, attack synthesis~\cite{Wen24ccs} or 
abstracting from actual contract code~\cite{Tolmach21wtsc,SunLSJ21wtsc,Babel23clockwork,Nielsen23cpp,Pusceddu24fmbc}.

\bibstyle{plainurl} 
\bibliography{main}



\end{document}